\documentclass[letterpaper, 10 pt, conference]{ieeeconf}  

\IEEEoverridecommandlockouts                              

\usepackage{graphicx} 
\usepackage{hyperref}
\hypersetup{
    colorlinks=true,
    filecolor=magenta,      
    urlcolor=blue,
}
\def\dotfill#1{\cleaders\hbox to #1{.}\hfill}
\makeatletter
\def\myrulefill{\leavevmode\leaders\hrule height .7ex width .2ex depth -0.6ex\hfill\kern\z@}
\makeatother
\usepackage{array}
\usepackage{ragged2e}

\newlength{\rlength}
\newcommand{\ruletext}[2][\rlength]{%
  \noindent%
  \parbox{#1}{%
    \noindent\hrulefill\raisebox{-.3\ht\strutbox}{#2}\hrulefill\par}%
}
\usepackage{subcaption}
\usepackage{booktabs}
\usepackage{amsmath}
\usepackage{multirow}

\usepackage{verbatim}
\usepackage[utf8]{inputenc}
\usepackage{dtk-logos}
\usepackage{tikz}
\usetikzlibrary{mindmap,trees}
\usepackage{varwidth}
\usepackage{adjustbox}

\definecolor{mygray}{HTML}{ECEDED}
\definecolor{myorange}{HTML}{FFA384}
\definecolor{mygreen}{HTML}{CCE7CF}
\definecolor{myblue}{HTML}{C2E8F7}
\definecolor{myyellow}{HTML}{F2F4C1}

\title{\LARGE \bf Urban Sound Classification : striving towards a fair comparison }

\author{Augustin Arnault $^{0}$ \\ \href{mailto:augustin.a@free.fr}{augustin.a@free.fr} 
   \and Baptiste Hanssens \\ \href{mailto:hanssens@multitel.be}{hanssens@multitel.be} 
   \and Nicolas Riche \\ \href{mailto:riche@multitel.be}{riche@multitel.be} 
\thanks{$^{0}$Work done as part of Multitel internship}%
}

\begin{document}

\maketitle
\thispagestyle{empty}
\pagestyle{empty}

\begin{abstract}

Urban sound classification has been achieving remarkable progress and is still an active research area in audio pattern recognition. In particular, it allows to monitor the noise pollution, which becomes a growing concern for large cities. 

The contribution of this paper is two-fold. First, we present our DCASE 2020 task 5 winning  solution \cite{dcase2020_our} which aims at helping the monitoring of urban noise pollution. It achieves a macro-AUPRC of 0.82 / 0.62  for the coarse / fine classification on validation set. Moreover, it reaches accuracies of 89.7\% and 85.41\% respectively on ESC-50 and US8k datasets. 

Second, it is not easy to find a fair comparison and to reproduce the performance of existing models. Sometimes authors copy-pasting the results of the original papers which is not helping reproducibility. As a result, we provide a fair comparison by using the same input representation, metrics and optimizer to assess performances. We preserve data augmentation used by the original papers. We hope this framework could help evaluate new architectures in this field. 

For better reproducibility, the code is available on \href{https://github.com/multitel-ai/urban-sound-classification-and-comparison}{our GitHub repository} \cite{git_analysis}. \\

\end{abstract}


\section{INTRODUCTION} \label{INTRODUCTION}

Sound detection and classification is a broad field with numerous potential applications like music genre classification, bird song classification or emotion detection. The term sound covers a wide lexical field such as music, conversation, animal sound or even noise. This last type of sound has recently been gaining attention. Noise pollution is a growing concern for large cities such as New York City. It is one of the top most quality of life issues for urban residents in the United States as it can have repercussions on the health of citizens. Projects such as SONYC \footnote{Sounds of New York City} have emerged to study noise and find solutions to mitigate impacts on citizens.

Since 2010, deep learning has given a boost to research in a number of areas and quickly replaced classical machine learning algorithms by demonstrating superior performance on various tasks. Deep learning techniques, thanks in particular to the convolutional neural networks \cite{yann}, have emerged as a powerful strategy for learning feature representations directly from data. As a result, it requires a large amount of data. In order to meet this demand, DCASE \footnote{Detection and Classification of Acoustic Scenes and Events} challenges have provided lots of publicly available datasets and have gained an important research interest in audio pattern recognition since 2013.  This challenge also contributes to  research by allowing participants to face their continuously improving models. 

There is still a need for large scale datasets featuring generic real-world sound just like ImageNet in image classification or Wikipedia data in natural language processing. To address this lack of data, in 2017, Google released AudioSet \cite{audioset}. This dataset contains 2.1 millions of 10-second audio sounds grabbed from YouTube videos and annotated with a hierarchical ontology of 527 types of sound events. Other noise pollution datasets have emerged such as ESC-50, UrbanSound8k and more recently SONYC-UST in the DCASE 2019 / 2020 task 5 challenges.

The first contribution of this work includes our DCASE 2020 task 5 winning solution. The source code can be found on \href{https://github.com/multitel-ai/urban-sound-tagging}{our Github repository} \cite{git_dcase}. Our model out-performs all the competing architectures of the challenge. In this paper, we also want to investigate if it is the case with respect to other models or datasets in the field of urban sound classification. As a result, the second contribution consists of a serie of experiments using our solution and other state of the art methods on specific noise pollution datasets. Additionally, we tried to build an objective framework by using the same input representation, metrics and optimizer to assess the performance of selected architectures. 

This paper is organized as follows. Section \ref{RELATED_WORK} presents various works carried out in the audio pattern recognition field. Section \ref{PROPOSED_METHOD} details the architecture of the model we developped and how we pre-processed the inputs. Section \ref{EXPERIMENTS} is a comparison of our model versus other state of the art models. Finally, in Section \ref{DISCUSSION}, discussions and further work perspectives are given. We conclude in Section \ref{CONCLUSIONS}.

\section{RELATED WORK} \label{RELATED_WORK}

We divided the work done into 4 main topics. The first one is input representation, which gathers all the techniques in order to find the best model input. The second one is data augmentation, which summers up all the techniques used to artificially increase the size of the dataset and consequently improve generalization. The third one is related to the model architecture options. Finally, the last topic is about the training techniques.

\subsection{Input Representation}

For a long time, feeding neural network models with raw audio was not possible because of a too high spatio-temporal complexity \cite{yann}. Spectrograms were used to overcome this problem. Additional transformations are applied to take the psychoacoustic knowledge into account. One of the most popular transformations is the Mel frequency scale. The use of spectrograms still dominates the field, as it can be seen in \cite{panns}, \cite{depthwise_sep} and \cite{TFNet}.
As those transformations are biased by human perception, \cite{convRBM} tries to learn better transformations to achieve better accuracy. More recent architectures have come back to the use of raw audio input \cite{wavems}. Raw data allows for a more end to end approach and prevents human biased feature extraction. Finally, some solutions such as \cite{panns} use a hybrid approach by using both spectrogram and raw data input to further increase the accuracy of the model.

\subsection{Data Augmentation}

In order to solve overfitting problems and improve generalization, various data augmentations have been studied. Since there are several possible input representations (mostly raw audio or spectrogram), many data augmentation techniques have been developed. On raw audio, the usual augmentation techniques are time stretching, pitch shifting, dynamic range compression and background noise addition, as described in \cite{audioDA}. Spectrogram based approaches, interpreted as a grayscale image, apply many computer vision data augmentation techniques \cite{da}. SpecAugment \cite{spectaugment} uses time warping, frequency masking and time masking. Argus, the Freesound Audio Tagging 2019 winning solution \cite{argus}, uses random resizing in addition to the techniques used by \cite{spectaugment}. 

\subsection{Architecture}

\begin{figure*}[ht]%
    \centering
    \includegraphics[width=\textwidth]{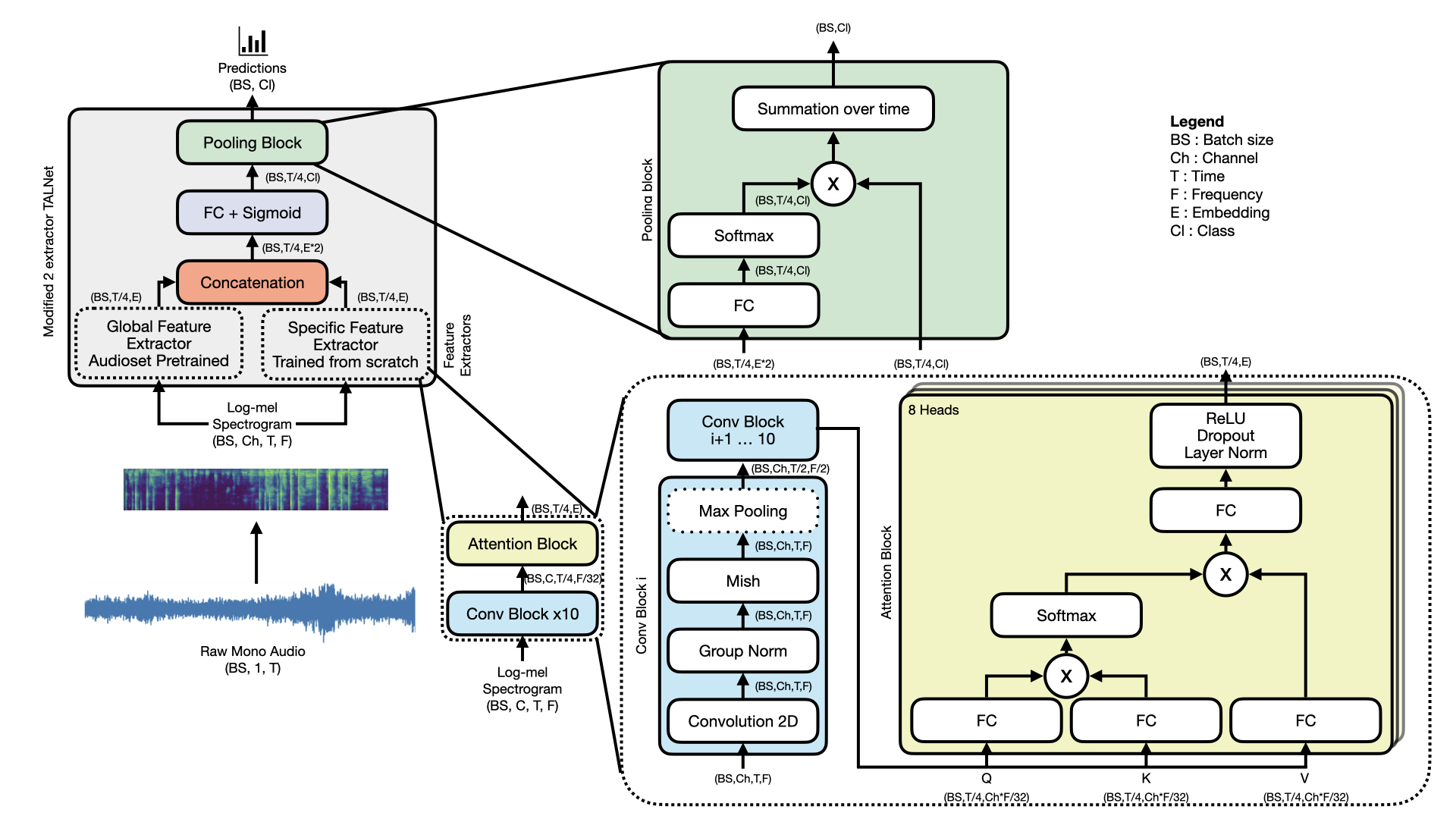}
    \caption{General architecture of our model. The raw audio is converted into a log-mel spectrogram. It passes into the global and specific feature extractor. Those extractors have 10 convolutionnal blocks (TALNet convolutions for the global feature extractor) and a RNN-like layer (Bi-GRU for the global feature extractor and multi-head attention for the specific feature extractor). Features are then concatenated and passed into a fully connected layer with sigmoid activation. At this state, we have probabilities of sound occurrence for each frames. The pooling layer gathers those probabilities per frame and outputs probabilities of sound occurrence for the sample.}
    \label{fig:archi_globale}%
\end{figure*}

Most audio classification architectures can be summarized with the following equation \ref{eq:general_archi}.

\begin{equation*}  \label{eq:general_archi}
\hat { Y } =({ f }_{ FC }\circ { f }_{ RNN }\circ { f }_{ CNN })({ X }_{ Input })
\tag{0}
\end{equation*}

$f_{FC}$ represents one or more fully connected layers added at the end of the model. If $f_{ RNN }$ is equal to the identity function ($Id$), the model is a pure CNN. In the most general case, $f_{ CNN }$ and $f_{ RNN }$ are both different from $Id$ and build a CRNN architecture. Those functions are obtained by training the model to minimize a loss function.

\ruletext[2.5cm]{ \textbf{CNNs} }
One of the first applications of CNNs in the audio field dates back to the early 90s with the work of Yann Lecun et al \cite{yann}. This work paved the way for the use of ANN \footnote{Artificial Neural Network} models that were previously inapplicable on this kind of data.

CNNs have been extensively studied and applied over the last decade. Several deep learning neural networks have been proposed in machine listening research. As mentioned and summarized in \cite{panns}, several CNN-based architectures have been applied to spectrogram of audio recordings followed by an activation function to predict the presence or absence of sounds.

As an example, the winning solution of  DCASE 2019 task 5 challenge used a modified pretrained version of MobileNetV2 \cite{dcase2019}. This model achieved better results than models focused directly on audio waveforms.

When the input representation is raw audio, the model is usually using 1D convolutions. Those layers are used in \cite{wavems}. For a given kernel length, 1D convolutions have less parameters than their 2D counterpart, assuming the kernel is a square matrix. Therefore, 1D convolutions are faster and take less memory than 2D convolutions.

CNNs do not have a very wide field of view and do not tackle long time patterns. To address this issue, \cite{depthwise_sep} used dilated convolutions and improved the accuracy.

\ruletext[2.5cm]{ \textbf{CRNNs} }
Although CNNs act as a robust feature extractor, their receptive fields have limited size and cannot capture long time dependencies. To solve this problem, CRNNs were proposed to consider the long time information. The idea is to use a CNN combined with RNN-like layers such as GRU or LSTM. \cite{talnet} proposed a CNN followed by a GRU layer. This model includes a final pooling layer as it was designed for MIL \footnote{Multiple Instance Learning} problems. Extensive comparison of pooling strategies have been made in \cite{talnet-pooling}.

As explained in \cite{cnn-transfo}, Transformers have been proposed to take the long time dependency of time series into account. This approach is inspired by the "Attention Is All You Need" work, well-known in the NLP  \footnote{Natural Language Processing} field \cite{attention}. More recent studies have shown that under certain conditions, transformer layers are like RNNs layers \cite{transformerRNN}. Replacing RNN layers by self attention layers has proven to increase performance in general.

\subsection{Training}

Transfer learning experienced a boom around 2016 with the proliferation of model zoos and tools to easily use this technique. It involves training the model on a huge dataset related to the same task. Once this training is done, the model weights are saved and loaded back to fine-tune a new model on a specific task and dataset. Models using transfer learning usually train quicker and achieve better performance : the model acquires a more general understanding of the task \cite{super-convergence}. However, it requires training the same or slightly modified model two times. Some techniques such as weight standardization are not compatible with pretrained models not using it \cite{GC}. Modification of the model between the pretraining phase and the transfer learning phase can cancel the benefits of transfer learning. Indeed, the previous learned representation is less relevant for another architecture. Literature revealed that ImageNet-pretrained models usually outperforms Audioset-pretrained models \cite{esresnet, dcase2019}.

\section{PROPOSED METHOD} \label{PROPOSED_METHOD}

In this section, we describe our solution taking log-mel spectrograms as input and label probabilities as output. We can split our contribution into 3 main points : data augmentation, neural network architecture and relabeling.

\subsection{Data augmentation}

We used SpecAugment \cite{spectaugment}, which  consists  of  warping  the features, masking blocks of frequency channels, and masking blocks of time steps to supplement the training data. Moreover, several image data augmentation techniques \cite{da} were used such as ShiftScaleRotate, Grid distortion and Cutout from the Albumentation library. We also tested Mixup \cite{mixup}, a method that linearly mixes two random training examples with a scalar lambda sampled from a beta distribution. We found out it was helping the model to obtain better scores when the model is trained from scratch. Nonetheless, as it hurt performance on our final solution, we deactivated it.

\subsection{Neural network architecture}\label{OUR_ARCHI}
The feature extractor of the TALNet model \cite{talnet} is the starting point of our final architecture. This extractor is a CRNN which consists of 10 convolutional layers with kernel of 3x3, interleaved with 5 max pooling layers, followed by one bidirectional GRU layer and one fully connected layer. Each max pooling layer reduces the number of frequency bins by half and the two first also reduce the frame rate by half. The output of the last max pooling layer is flattened and fed into the bidirectional GRU layer.  

For our solution, we opted for a dual backbone. One develops a global understanding of the sample while the other develops a specific understanding to the dataset used. The first feature extractor, as described above, is similar to the original TALNet-like feature extractor. We used the pretrained Audioset version so that it acts as a global extractor. The other feature extractor is our modified implementation of the previously mentioned TALNet layers and acts as a specific extractor. 

Four main improvements have been incorporated into the specific feature extractor. First, Group Normalization (GN) \cite{gn} is used instead of Batch Normalization (BN). GN divides the channels into groups and computes the mean and variance for normalization within each group. GN’s computation is independent of the batch size, and its accuracy is stable in a wide range of batch sizes. A second normalization technique called Gradient Centralization (GC) \cite{GC} is used to accelerate the training and smooth the loss and the gradients. It operates directly on gradients by centralizing the gradient vectors to have a zero mean. GC can be viewed as a projected gradient descent method with a constrained loss function. In our model, GC is only applied on convolution layers. Thirdly, we replaced the activation function by the Mish function \cite{mish}. It provides an overall lower loss, smoother and a well conditioned easy-to-optimize loss landscape. However, compared to ReLU, this is done at the price of higher memory usage and computational time. Last but not least, the bidirectional GRU layer has been replaced by a multi head self attention block. An attention function can be described as mapping a query and a set of key-value pairs to an output, where the query, keys, values and output are all vectors. Here query, keys and values are the same tensor (the concatenated features) because we want to use a self attention mechanism  \cite{attention}. This layer is a replacement of the bidirectional GRU, decreasing the number of parameters and increasing performance.  

The output of the dual backbone (feature extractors) is fed into a fully connected layer followed by a sigmoid activation function. At this state, we have probabilities for each frame and for each class. However, the classification task gives us record level labels. We use an attention pooling block to aggregate predictions of each frames. The architecture is summarized in figure \ref{fig:archi_globale}.

\subsection{Relabeling}
This section is specific to the SONYC-UST dataset in which the labels are raw crowdsourced. Volunteers are not experts and can mislabel a sample or miss a label, especially since the annotation is multi-label. As a result, three different noisy annotations are available for each sample but we had to deal with the reliability of the labels.

The default strategy relies on computing the mean of the three annotations for each sample and defining a threshold to get a more reliable label. On top of that, we applied a relabeling strategy to annotate all the training set. The $\sim$500 samples annotated by SONYCUST team remain untouched. 

We first train a model with the dataset annotated by three different volunteers and the default strategy. The training was stopped after reaching the highest macro AUPRC on coarse-level labels. This model is then used to relabel samples. Third, a new model is trained on the relabeled dataset. We found out this strategy can help when the metric to optimize is high enough. As shown in Table \ref{tab:results}, we think it was one of the decisive tricks.

 \section{EXPERIMENTS} \label{EXPERIMENTS}

\begin{table*}[ht]
\begin{center}
\caption{Results of experiments}
\label{us8k_perf}
\begin{tabular}{lll|c|cc|c|c}
\toprule
 & & & DWSDC & \multicolumn{2}{c|}{CNN10} & TALNet modified & TFNet \\
  & & & from scratch & from scratch & pretrained audioset& pretrained audioset (+Relabel) & from scratch \\\midrule
\multirow{8}{*}{\rotatebox[origin=c]{90}{SONYC-UST}} & \multirow{4}{*}{\rotatebox[origin=c]{90}{Coarse}} & $F1_{micro}$ &0.776 &0.792 &0.798 &0.808 (\textbf{0.819}) &0.785\\
 & & $AUPRC_{micro}$ &0.865 &0.884 &0.890 &0.892 (\textbf{0.898}) &0.875\\
 & & \textbf{AUPRC$_{macro}$} &0.757 &0.777 &\textbf{0.823} &0.806 (0.820) &0.702\\ 
 & & $mAP$ &0.759 &0.782 &\textbf{0.825} &0.810 (0.823) &0.708\\
 \cmidrule(lr){2-8}
 & \multirow{4}{*}{\rotatebox[origin=c]{90}{Fine}} & $F1_{micro}$ &0.683 &0.717 &0.722 &\textbf{0.723} (\textbf{0.723}) &0.706\\
 & & $AUPRC_{micro}$ &0.751 &0.794 &0.802 &0.798 (\textbf{0.808}) &0.779\\
 & & \textbf{AUPRC$_{macro}$} &0.485 &0.579 &0.559 &0.585 (\textbf{0.623}) &0.530\\ 
 & & $mAP$ &0.508 &0.597 &0.576 &0.604 (\textbf{0.637}) &0.547\\
\midrule
\multirow{5}{*}{\rotatebox[origin=c]{90}{US8K}}& \multirow{5}{*}{\rotatebox[origin=c]{90}{Mono}}& \textbf{Accuracy} &0.793$\pm$0.046 &0.838$\pm$0.050 &\textbf{0.861$\pm$0.039} &0.854$\pm$0.040 &0.824$\pm$0.052\\
 & & $F1_{micro}$ &0.783$\pm$0.043 &0.836$\pm$0.049 &\textbf{0.860$\pm$0.039} &0.852$\pm$0.039 &0.825$\pm$0.048\\
 & & $AUPRC_{micro}$ &0.816$\pm$0.061 &0.897$\pm$0.039 &\textbf{0.927$\pm$0.026} &0.922$\pm$0.026 &0.887$\pm$0.043\\
 & & $AUPRC_{macro}$ &0.859$\pm$0.056 &0.908$\pm$0.035 &\textbf{0.929$\pm$0.028} &0.923$\pm$0.028 &0.898$\pm$0.039\\ 
 & & $mAP$ &0.858$\pm$0.058 &0.908$\pm$0.034 &\textbf{0.930$\pm$0.028} &0.924$\pm$0.028 &0.898$\pm$0.039\\
\midrule
\multirow{5}{*}{\rotatebox[origin=c]{90}{ESC-50}}& & \textbf{Accuracy} &0.798$\pm$0.032 & 0.869$\pm$0.022&\textbf{0.900$\pm$0.0170} &0.897$\pm$0.022 &0.841$\pm$0.016\\
 & & $F1_{micro}$ &0.775$\pm$0.031 & 0.855$\pm$0.023 &\textbf{0.890$\pm$0.013} &0.871$\pm$0.017 &0.798$\pm$0.015\\
 & & $AUPRC_{micro}$&0.794$\pm$0.040 &0.918$\pm$0.020 &\textbf{0.950$\pm$0.009} &0.940$\pm$0.014 &0.902$\pm$0.015\\
 & & $AUPRC_{macro}$ &0.821$\pm$0.034 & 0.923$\pm$0.021 &\textbf{0.944$\pm$0.010} &\textbf{0.944$\pm$0.015} &0.908$\pm$0.016\\ 
 & & $mAP$ &0.829$\pm$0.032 & 0.927$\pm$0.019 &\textbf{0.947$\pm$0.010} &\textbf{0.947$\pm$0.014} &0.913$\pm$0.015\\
\bottomrule
\label{tab:results}
\end{tabular}
\caption*{Metrics in bold font are the official one used for ranking. For US8K and ESC-50, the result is the mean of the highest score reached during training over each fold $\pm$ the standard deviation.}
\end{center}
\end{table*}

It was not easy to find a framework that could be used to evaluate our model. Moreover, most of the time, authors just copy paste the results from the original paper which does not encourage repoducibility. As a result, we tried to build a fair comparison described in this section. In order to do that, we used the same input representation, metrics and optimizer to assess the performance of selected architectures.

\subsection{Input Representation}

\begin{table}[ht]
\begin{center}
\caption{Hyper-parameters used for log-mel spectrograms}
\label{hp_spec}
\begin{tabular}{llllll}
\toprule
$n_{fft}$ & $window$ & $hop length$ & $n_{mels}$ & $f_{min}$ & $f_{max}$ \\\midrule
2822 & Hanning & 1103 & 64 & 0 & 8000\\
\bottomrule
\end{tabular}
\end{center}
\end{table}

For a fairer comparison, the input representation is the same for all experiments, hence all models have access to the same quantity of information as input. Recordings are first resampled at 44100 Hz then transformed into log-mel spectrograms. This operation is done on CPU using Librosa \cite{librosa} and appears to be the main bottleneck. We decided to save all the log-mel spectrograms on the hard drive to avoid the same computation at each epoch of the training. The main hyper-parameters used to compute the log-mel spectrograms (STFT and mel-scale transform) are summarized in table \ref{hp_spec}.

Data augmentation is used when specified in the tested architecture original paper. We decided to do so because some data augmentation could hurt performance when applied on a specific architecture (as revealed when testing mixup on our dual extractor architecture).

\subsection{Datasets}

For our comparison, we used 3 different datasets cited in Section \ref{INTRODUCTION}. Those datasets have specificities mentioned in table \ref{tab:overview_data}.

\begin{table}[ht]
\begin{center}
\caption{Overview of audio datasets}
\label{overview_dataset}
\begin{tabular}{l|llll}
\toprule
Dataset & Total duration & Classes & Length & Channels \\\midrule
Audioset & 241 d & 527 & $\leq$10s & Mono\\
SONYC-UST & 51.4 h & 8 / 23 & 10s & Mono\\
UrbanSound8k & 8.75 h & 10 & $\leq$4s & Stereo\\
ESC-50 & 2.8h & 50 & 10s & Mono\\
\bottomrule
\end{tabular}
\caption*{The table above summarizes the principal characteristics of mainstream audio datasets for classification tasks}
\label{tab:overview_data}
\end{center}
\end{table}

\ruletext[2.5cm]{ \textbf{SONYC-UST} } The task 5 of the DCASE challenge provides SONYC-UST \footnote{SONYC Urban Sound Tagging} \cite{sonycust} as main dataset. It provides about 17,000 10-second samples with coarse-grained and fine-grained tags alongside a diversity of metadata such as spatio-temporal context. This dataset has been recorded by the SONYC acoustic sensor network and tagged by volunteers and SONYC team members. Samples can be polyphonic, meaning it can contain multiple sounds that can overlap. It is probably today's best dataset to test models for real-world noise classification.
This dataset has a given predefined training validation split. For the validation set, we only use the ground truth provided by the SONYC expert team. All of the 10-second samples are converted into log-mel spectrograms (400 frames, 64 mel bins). We used our relabeling method on our model only because we realised that the quality of training set annotations was not that good.

\ruletext[2.5cm]{ \textbf{US8K} } The UrbanSound8k dataset provides another good benchmark. It contains 8,732 labeled sound excerpts of urban sounds from 10 classes. All samples are taken from www.freesound.org and are monophonic but they have variable lengths ($\leq$4s) \cite{us8k}. This dataset has 10 predefined folds to use for cross validation. 
We use the cross validation over the 10 predefined folds. Samples being stereo, we could access even more information by providing the 2 channels. However, to keep things simple, we converted everything into a single channel. Additionally, samples with variable sizes are zero-padded to obtain a certain fix length for all samples. They are then converted into log-mel spectrograms (162 frames, 64 mel bins).

\ruletext[2.5cm]{ \textbf{ESC-50} } The ESC-50 dataset is a labeled collection of 2,000 environmental audio recordings from www.freesound.org suitable for benchmarking methods of environmental sound classification. Since UrbanSound8k and ESC-50 are extracted from FreeSound, there is a risk of data redundancy when used at the same time. This risk also exists when using Audioset and a FreeSound based dataset. The dataset consists of 5-second-long recordings organized into 50 semantical classes (with 40 examples per class) loosely arranged into 5 major categories \cite{esc50}. All samples are monophonic, meaning there is only one sound per sample. To ensure a fairer comparison between models, as proposed in this dataset, the evaluation is done with cross-validation over 5 predefined folds.
We use cross-validation over the 5 predefined folds. All classes are balanced and the folds are stratified. The samples have fixed lengths and are converted into log-mel spectrograms (200 frames, 64 mel bins).

\subsection{Selected architectures}

We selected the best models on the selected datasets in addition to our model. To ensure a fairer comparison, we used
our optimizer based on RangerLAMB \cite{RangerLars} applying GC on convolutionnal layers only. Each model has been integrated in our training loop using Pytorch Lightning \cite{pl}.

First, we selected a Depthwise Separable and Dilated Convolutionnal neural network (DWSDC) \cite{depthwise_sep}. Although this model was intended to be used for Sound Event Detection (SED) like TALNet, we added our pooling layer at the end of the network. We used the best performing architecture of the paper with inner kernel size of 3 and inner padding of 1 and no data augmentation. The batch size is 24.

TFNet was claiming really high accuracy on US8K in comparison to other models \cite{TFNet}. It features an interesting attention mechanism along the time and frequencies we wanted to compare to our model. We adapted the kernel size of pooling layers to fit our input representation. Batch size is 42.

CNN10 from the PANNs paper \cite{panns} was the final model considered. It is the state of the art on many audio datasets. For this model, we used Mixup with $\alpha=1$, Specaugmenter, a batch size of 64 and we removed the part computing log-mel spectrograms.

We also considered training models from scratch and pretrained counterpart, again to ensure a fairer comparison. We did not train our model from scratch because this architecture was designed to have a pretrained part : the global feature extractor described in section \ref{OUR_ARCHI}.

\subsection{Evaluation metrics}

For our experiments, we have chosen common metrics to compare the performance of the selected models. Those metrics are computed by the scikit-learn \cite{sklearn} metric module or an extended version of this module. Each dataset uses a specific metric for evaluation, they are in bold in table \ref{tab:results}. The threshold used for rounding predictions is 0.5.

The first selected metric is accuracy. It is easy to understand and is generally a good starting point to assess general performance. However, it does not take data distribution into account and fail to provide useful information when the dataset is not balanced.

The second metric selected is F1 score or harmonic mean of the precision and recall. This metric is more suited for unbalanced dataset than accuracy. However, it gives equal importance to precision and accuracy, which may not be ideal in particular cases. 

We also picked the mean Average Precision (mAP) and macro Area Under Precision Recall Curve (macro AUPRC). It summarizes the precision-recall curve as a single score. These metrics are equivalent and should output the same results. However, the mAP historically computed by averaging the precision over a set of evenly spaced recall levels (usually 11) may be different from the macro AUPRC. Here, mAP is computed by taking each unique point of the precision recall curve. We will show that we obtain similar results to macro AUPRC with the exception of a few trials. 

Finally, we selected the micro AUPRC. It is obtained by applying AUPRC over the flattened one hot predictions, as in a binary classification task.

\subsection{Analysis}

The results of our experiments are shown in table \ref{tab:results}. In order to be complete, all metrics have been used for each dataset. The official metric for ranking is showed in bold font. Moreover, for USK8K and ESC-50, we used the cross validation proposed by their authors and the result is the mean of the highest score reached during training over each fold and the standard deviation. To ensure an fair comparison, we compared the models separately. On one hand trained from scratch and on the other hand using the pretrained models.

For the trained from scratch approaches, we compared DWSDC, TFNet and CNN10. DWSDC model performs the worse with accuracies of 79.8\% and 79.3\% respectively on ESC-50 and US8k datasets. However it was not originally suited for this task and it may be possible to find a better pooling head for this architecture. TFNet achieved 84.15\% in accuracy on ESC-50, which is much lower than what they announced in their paper (87.70\%) \cite{TFNet}. In the ESResNet paper, they found an even lower accuracy (79.45\% with no augmentation policy). We speculate, like ESResNet \cite{esresnet} author, the dataset was misused : predefined folds were shuffled and split. CNN10 is the best model among models trained from scratch.

For pretrained models, we compared the 2 best solutions for DCASE 2020 task 5 : CNN10 and our modified TALNet. CNN10 pretrained on audioset matches the performance of the modified TALNet model on the SONYC-UST dataset except on fine-level categories. We believe that the multi head self attention layer allows to grasp longer time dependent pattern than convolutions. CNN10 outperforms our solution on datasets having a good quality of annotation. We found that our relabel strategy has played a determining role on SONYC-UST. The relabel strategy was used because we found out that labels were not accurate enough especially on fine-grained labels. For instance, some examples have been labeled as belonging to each class. This dataset corruption is due to the labeling system : volunteers tagged the audio clips on internet. They do not have the same knowledge as an expert and even experts do not always agree. A consensus among the SONYC team annotations is given for some samples.

\section{DISCUSSION} \label{DISCUSSION}

In this section, we will raise many issues found in the audio pattern recognition field. Hints and further work perspectives will be given.

\subsection{Computing spectrograms}
In order to compute spectrograms, we experimented with frameworks like Torchaudio, Librosa and even with our own implementation in Pytorch. Pytorch based approaches allowed a x2 trade-off in time duration because everything was computed on GPU. However, we found a drop in performance when not using Librosa and we did not manage to find why performance was lower. The way log-mel spectrograms are computed can introduce a bias when a model is using one implementation doing the transform differently. One way to address this issue is to give a precomputed log-mel spectrogram as used in TUT-SED Synthetic 2016 dataset \cite{tutsed}. Moreover, it is not clear whether resampling before the spectrogram transform is helping or not. On US8K datasets, it is often not mentioned whether audio was used in mono or stereo mode. We think the domain lack a study comparing the different ways to compute log-mel spectrograms and their related hyperparameters impact.

\subsection{Computing metrics}
We found discrepancies in obtained metrics. The source of bias is the framework computing the metrics. For instance, the implementation of area under curve (AUC) is computed differently from one framework to another. We found out discrepancies on metrics like F1 score. We chose scikit-learn to compute everything because the framework is older than pytorch (and pytorch lightning) and well tested. Even within the same framework, we observed disparities between mAP and macro AUPRC because of the way it is computed. The order of operations were not the same.
In that way, we encourage reproducibility whenever possible.

\subsection{Model Comparison}

We realised it was not easy to find a good benchmark, even if ESResNet paper \cite{esresnet} paved the way for a better comparison. The authors highlight the misuse of the predefined folds, leading to results higher than other approaches. We also had a hard time figuring out whether cross validation was used in some papers. Most of the time, standard deviation was not mentioned and there was only the mean score.

The code is now crucial and must be released because some techniques have different variants. For instance, we found at least 3 ways to do Mixup and we were not able to find which implementation was the best. We also tried to replicate a MobileNetV2 approach without success. The model was performing worse than the selected model. More investigation is needed.

\section{CONCLUSION AND FURTHER WORK} \label{CONCLUSIONS}

In this paper, a new architecture for Urban Sound Classification has been proposed. Our model outperforms existing methods on the DCASE2020 Task5 challenge and is ranked among the top models on other datasets (ESC-50 and US8k).

To evaluate our solution, we introduce an impartial framework that uses the same input representation, metrics and optimizer in order to assess the performance of the architectures only. We hope this framework will help evaluate new algorithms in urban sound classification.

In the future, we would like to inlcude ImageNet-based models such as ESResnet \cite{esresnet} or MobileNet \cite{dcase2019}. A class-wise analysis could also be performed in order to better understand which approach is more efficient at detecting a specific set of sound. A study about model parameters or inference time could also be carried out. We would like to train a new modified TALNet using transfer learning only with one fine-tuned feature extractor and Gradient Centralization (GC).

\section{ACKNOWLEDGEMENT}
This research was carried out as part of the Wal-e-Cities project portfolio, co-financed the European Regional Development Fund (ERDF 2014-2020) and supported by the Walloon Government.



\end{document}